# CALCULATIONS OF AVERAGE NUMBERS OF PROMPT NEUTRONS FOR ACTINIDE PHOTOFISSION

**A.I. Lengyel[1), O.O. Parlag[1], V.T. Maslyuk[1], N.I. Romanyuk[1], O.O. Gritzay[2]**

*[1] Institute of Electron Physics*
*Universitetska 21, Uzhgorod 88017 Ukraine, parlag.oleg@gmail.com*
*[2] Institute for Nuclear Research*
*Prospekt Nauky 47, Kiev 03680 Ukraine*

The empirical calculations of the prompt neutrons average number $\bar{\nu}$ for photofission of the $^{232}$Th, $^{233}$U, $^{234}$U, $^{235}$U, $^{236}$U, $^{238}$U, $^{237}$Np, $^{239}$Pu and $^{241}$Am actinides has been done as a function of photon energy, mass and charge of the nuclei, which can be used to evaluate the $\bar{\nu}$ for photo-fission of nuclides for which no or scarce data are available.

The average number of prompt neutrons is one of two (along with the fission cross section) major nuclear-physical parameters required for practical calculations. This value is determined in detail and accurately for neutron-induced reactions for the most nuclides. At the same time, the experimental data and evaluation of the average number of prompt neutrons in the case of photofission are much scarcer, old-fashioned or obtained by indirect mode [1,2]. With the increasing interest in the methods of nuclear fuel burning and long-lived actinide decontamination the need in precise values of nuclear constants is evident during last years. This is especially true for photonuclear constants. Therefore, the search was focused on a formula that can be used to estimate the value of $\bar{\nu}$ for photofission of arbitrary actinide.

Let's consider a general approach to modeling the average number of the prompt neutrons in the case of neutron activated fission. This value is a function of three variables: the mass of the compound nucleus *A*, its charge *Z* and the incident neutron energy $E_n$ ($\bar{\nu} \equiv \bar{\nu}(A,Z,E_n)$).

In previous studies [3-5] it has been shown that expanding the charge, mass and energy dependence of $\bar{\nu}$ in the form of truncated Taylor series apparently yields a reasonable representation for $\bar{\nu}(A,Z,E_n)$ if the zero-, first- and one second-order cross term are kept in truncation for all isotopes, i. e. it is sufficient to use a linear approximation over all three variables, at least up to the threshold *(γ,2nf)*, taking into account the contribution of the even-odd effect in a common form:

$$\bar{\nu}(A,Z,E_n) = \bar{\nu}_0(A,Z) + a(A,Z)(E_n - E_0), \quad (1)$$

where the intercept

$$\bar{\nu}_0(A,Z) = C_1 + C_2(Z - Z_0) + C_3(A - A_0) + C_4 P(A,Z), \quad (2)$$

and slope

$$a(A,Z) = C_5 + C_6(Z - Z_0) + C_7(A - A_0) + C_8 P(A,Z), \quad (3)$$

$$P(A,Z) = 2 - (-1)^{A-Z} - (-1)^Z, \quad (4)$$

$Z_0$, $A_0$ and $E_0$ are the values, about which the expansion is to be made [3].

The coefficients $C_i$ in equations (1) – (3) were evaluated by the least-square method [5]. A wide range of existing data for $\bar{\nu}(A,Z,E_n)$ for actinides from $^{232}$Th to $^{245}$Cm was analyzed and the best simplified parameterization of the slope and intercept was recently found in [6]:

$$\bar{\nu}(A,Z,E_n) = \bar{\nu}_0(Z) + a(A)E_n, \quad (5)$$

$$\bar{\nu}_0(Z) = -22{,}7734 + 0{,}27318\,Z, \quad (6)$$

$$A(A) = -0{,}1636 + 0{,}0013\,A. \quad (7)$$

These calculations were performed in the case of neutron-induced actinide fission without taking into account other reactions.

Generally speaking, all mentioned approaches have to be versatile and suitable for all other cases of fission, including photofission. Therefore, we applied the model (1-4) in the case of photofission, where $(E_n - E_0)$ is replaced by $(E_\gamma - E_s)$, $E_\gamma$ – photon energy, $E_s$ – nucleon separation energy.

To do this, we take into account the wide experimental data set of photofission $\bar{\nu}(A,Z,E_\gamma)$ in the energy range ~ 8 – 20 MeV for actinides $^{232}$Th [7,8], and ~ 5 – 20 MeV for $^{233}$U [9,10], $^{234}$U [9,10], $^{235}$U [7,9], $^{236}$U [7,9], $^{238}$U [7,9], $^{237}$Np [9,10] and $^{239}$Pu [9,10]. We used all these 219 experimental points for our calculation.

The values of the nucleon separation energy for each actinide were taken from [11,12] and are presented in Table 1.

Table 1. Nucleon separation energy

| Actinide number | A | Z | $E_s$, MeV | References |
|---|---|---|---|---|
| 1 | 232 | 90 | 6.438 | [11] |
| 2 | 233 | 92 | 5.760 | [11] |
| 3 | 234 | 92 | 6.844 | [11] |
| 4 | 235 | 92 | 5.298 | [11] |

| 5 | 236 | 92 | 6.545 | [11] |
| 6 | 237 | 93 | 6.580 | [11] |
| 7 | 238 | 92 | 6.152 | [11] |
| 8 | 239 | 94 | 5.647 | [11] |
| 9 | 240 | 94 | 6.534 | [11] |
| 10 | 242 | 95 | 6.641 | [12] |

To evaluate the quality of linear approximation (1) – (4), modified for photofission, we calculate $\bar{v}$ for each actinide separately. The best possible energy approximation for $\bar{v}$ have to be lie within the linear approach [7,10]:

$$\bar{v}(E_\gamma) = \sum_{j=1}^{N} \sum_{i=1}^{j} f_i(E_\gamma) \theta(n), \quad (8)$$

$i, j$ – actinide number, $n = i - j$, $N = 8$, $\theta(n)$ - step function

$$\theta(n) = \begin{cases} 0 & n<0 \\ 1 & n \geq 0 \end{cases},$$

$$f_i(E_\gamma) = x_i + y_i(E_\gamma - E_s), \quad \text{for} \quad E_i^{min} \leq E_\gamma \leq E_i^{max},$$

where $E_i^{min}$ and $E_i^{max}$ are the lower and upper bound of photon energy interval for respective $i$-th actinide, $E_\gamma - E_s > 0$.

The common quality of the description of the data set for each actinide by (8) is evaluated by [13]:

$$\langle \chi^2 / dof \rangle = \frac{\sum_{i=1}^{N} \frac{\chi_i^2}{n_{data_i} - m_{par}}}{N}, \quad (9)$$

where each $i$-th interval out of total of N intervals contains $n$ data points and two parameters give a resulting combination to $\chi_i^2$ in fitting eq. (8) with result $\langle \chi^2/dof \rangle = 2.1$ (dot line in Fig. 1.). Consequently it is naturally that we have to build the most general formula describing the average prompt neutrons for photofission depending on the photon energy of all actinides, which is close to the best results of local linear calculation. As the initial formula, we chose that similar to (1):

$$\bar{v}(A,Z,E_\gamma) = \bar{v}_0(A,Z) + a(A,Z)(E_\gamma - E_s(A,Z)), \quad (10)$$

where the slope $\bar{v}_0(A,Z)$ and the intercept $a(A,Z)$ are chosen as (2) – (3).

The results are given in Fig. 1 (solid line). Coefficients $C_i$ were calculated by least-square method [14]. The final formula for calculating the average number of prompt neutrons for photofission of actinides is:

$$\bar{\nu}_0(A,Z) = (1{,}97 \pm 0{,}05) + (0{,}165 \pm 0{,}028)(Z - 90) +$$
$$+ (0{,}0341 \pm 0{,}0093)(A - 232) - (0{,}0853 \pm 0{,}0094) \cdot P(A,Z), \qquad (11)$$

$$a(A, Z) = (0{,}0963 \pm 0{,}75 \cdot 10^{-2}) + (0{,}0371 \pm 0{,}43 \cdot 10^{-2})(Z - 90) - \qquad (12)$$
$$- (0{,}566 \pm 0{,}138) \cdot 10^{-2} \cdot (A - 232).$$

The quality of the description of the experimental data in this case: $\chi^2/dof = 2.5$.

This value is in accord with the best linear approximation for each actinide.

We also checked the quality of the recommended values from ENDF/B-VII-1 [15] (dash lines in Fig. 1.) for the same data set. All the approaches are in accord with each other at least up to the threshold *(γ,2nf)*.

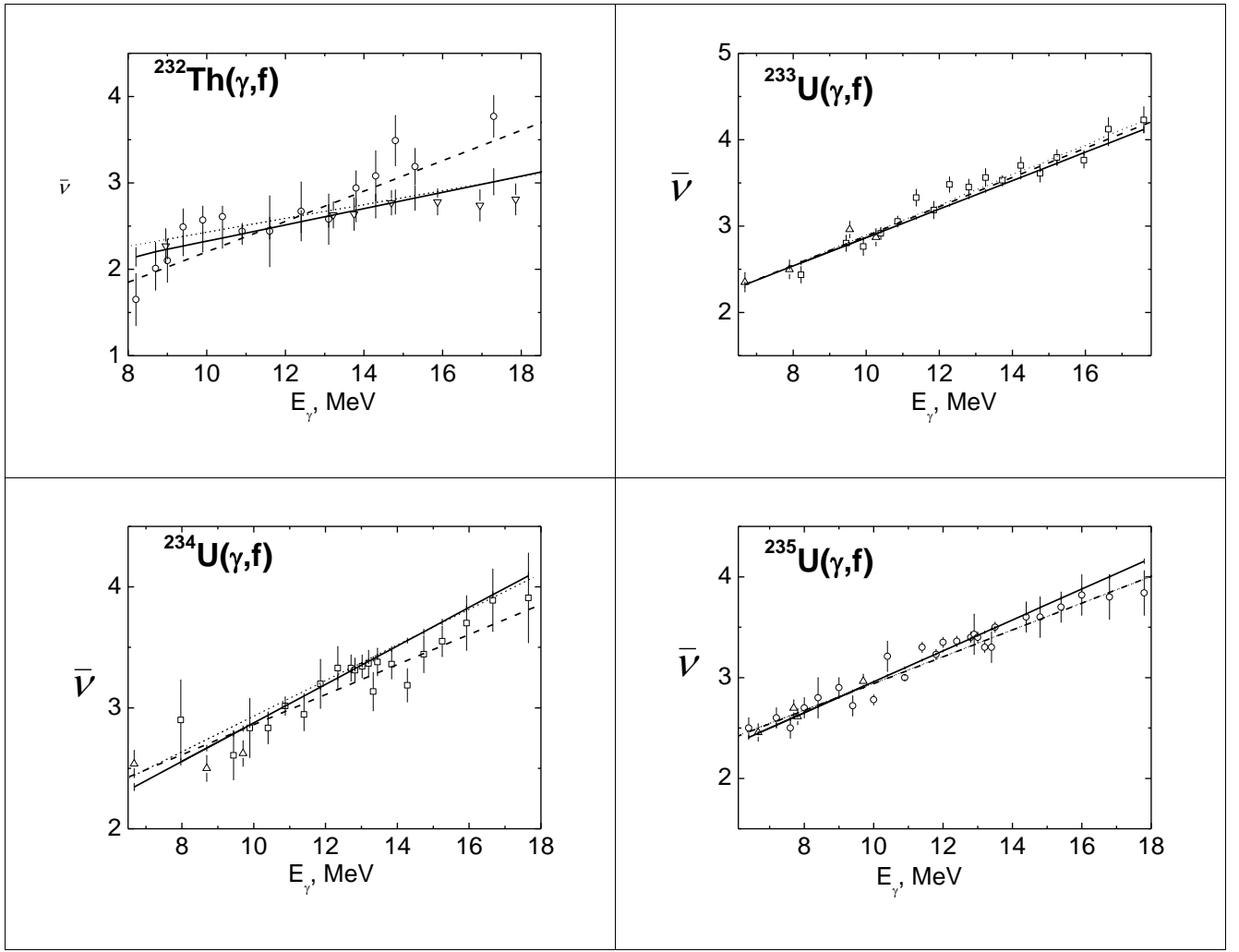

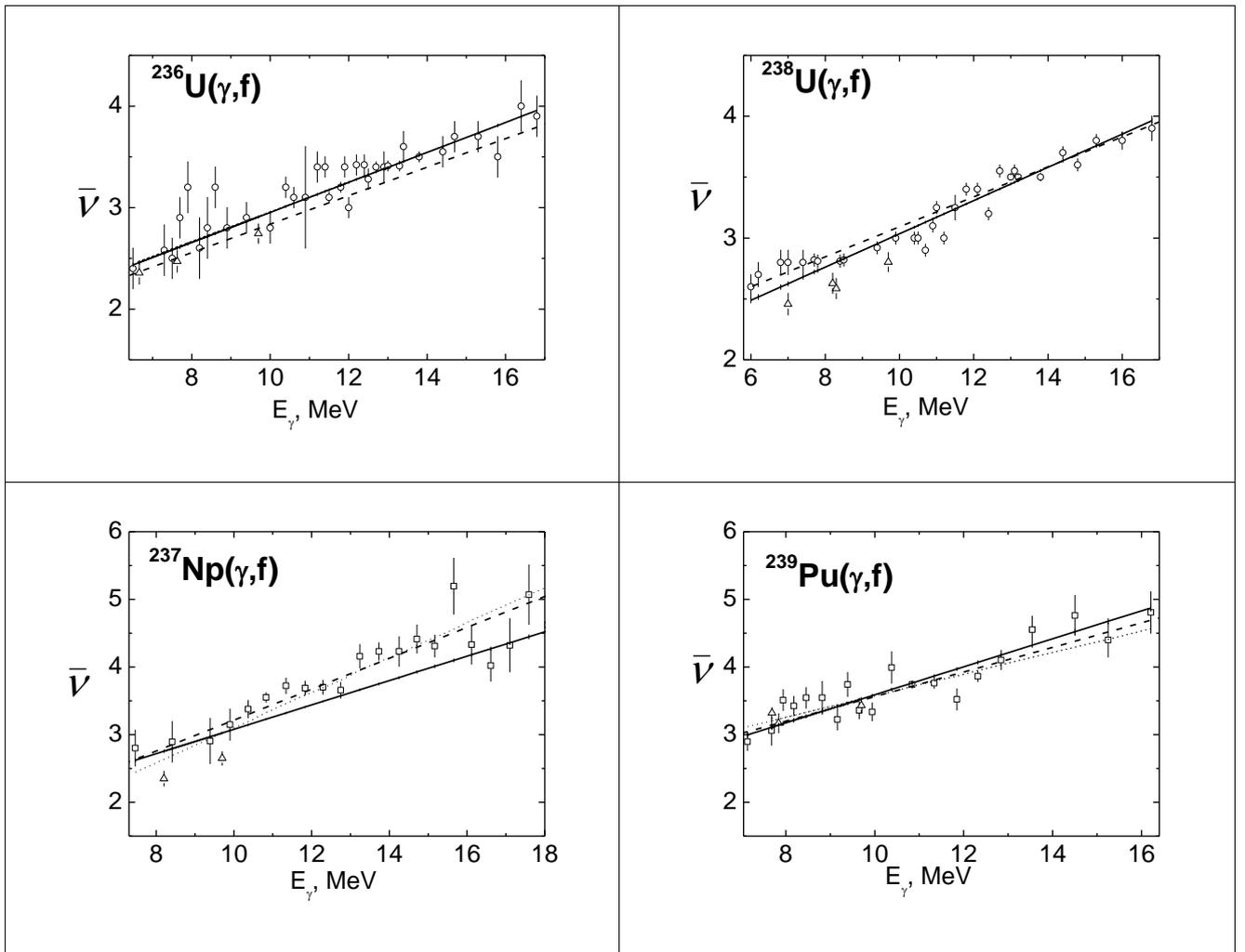

Fig. 1. The results of fitting experimental data for average neutron multiplicities using (10)-(12) (solid lines) for $^{232}$Th, $^{233}$U, $^{234}$U, $^{235}$U, $^{236}$U, $^{238}$U, $^{237}$Np and $^{239}$Pu: open circles (○), down triangles (▽), up triangles (Δ), and squares (□) – experimental data from [7,8,9,10] respectively; dash lines – ENDF/B-VII.1 [14]. Dot line – formula (8).

Also using formulas (10) – (12) we have calculated the photon energy dependence of $\bar{\nu}$ in the case of the $^{240}$Pu and $^{241}$Am [16] isotopes (see Fig. 2).

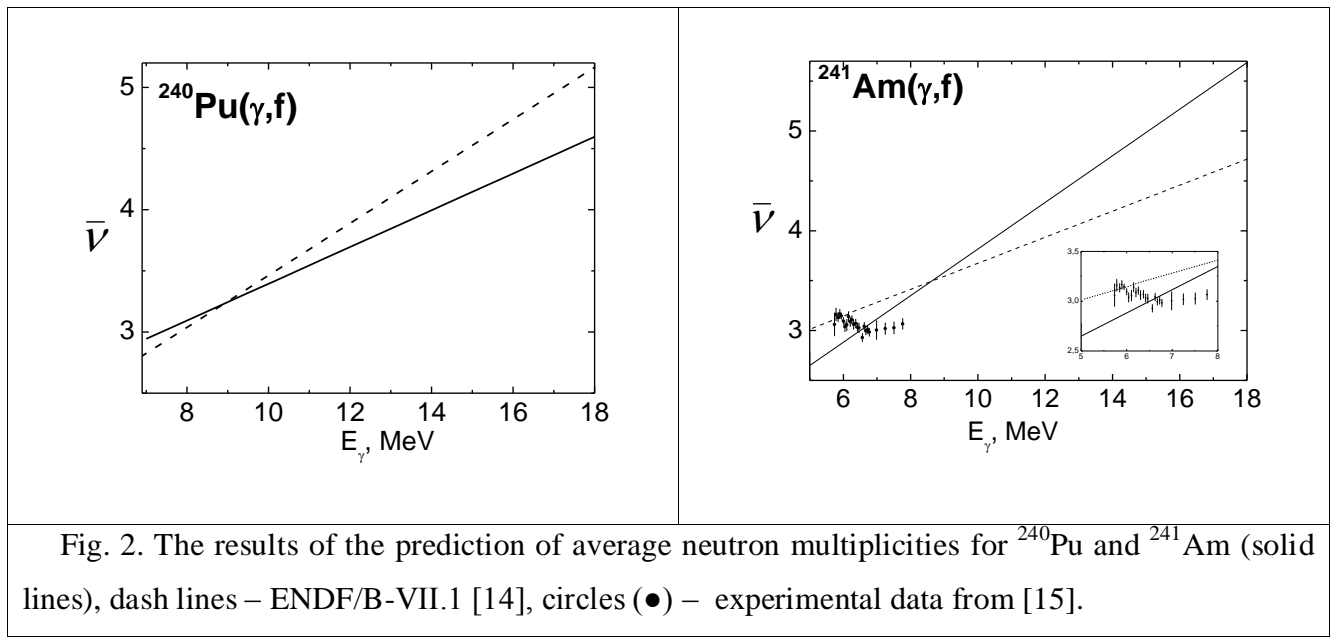

Fig. 2. The results of the prediction of average neutron multiplicities for $^{240}$Pu and $^{241}$Am (solid lines), dash lines – ENDF/B-VII.1 [14], circles (●) – experimental data from [15].

## Conclusion

The empirical formula for estimation of the average number of prompt neutrons $\bar{\nu}$ as a linear function of the $(E_\gamma - E_s(A,Z))$ of actinides photofission was obtained. The formula contains 7 free parameters taking into account the even-odd effect. The quality of description of $\bar{\nu}$ by formulae (10) – (12) ($\chi^2/dof = 2.5$) is comparable (see Fig.1.) with that of a local description for each actinide separately ($\chi^2/dof = 2.1$). All this gives us enough reason to use the resulting formula to calculate the average number of prompt neutrons $\bar{\nu}$ for any actinide photofission.

## References

1. *Giacri M.-L.* Photofission neutron emission. In: Development of photonuclear cross section library for CINDER'90 // http://giacri.free.fr/Articles/rapport_Los_Alamos.pdf

2. *Chadwick M.B., Oblozinsky P., Herman M. et al.* ENDF/B-VII.0: Next Generation Evaluated Nuclear Data Library for Nuclear Science and Technology // Nuclear Data Sheets – 2006. – V. 107. – P. 2931-3060.

3. *Howerton R.J.* $\bar{\nu}$ revised // Nuclear Science and Engineering. – 1977. – V. 62. - P. 438- 454.

4. *Bois R., Frehaut J.* Evaluation semi-empirique de $\bar{\nu}_P$ pour la fission induite par neutrons rapides // Commissariats a l'Energie Atomique report CEA-R-4791. (1976). – P.44-51

5. *Wahl A.C.* Phenomenological model for fragment mass and charge distribution in actinide nuclei fission // In: Fission product yield data for the transmutation of minor actinide nuclear waste. Vienna. 2008, p. 117-148.

6. *Ohsawa T.* Empirical formulas for estimation of fission prompt neutron multiplicity for actinide nuclides // Journal of Nuclear and Radiochemical Sciences. – 2008. – V. 9, No.1. – P. 19-25.

7. *Caldwell J.T., Alvarez R.H., Berman B.L. et al.* Experimental determination of photofission neutron multiplicities for $^{235}$U, $^{236}$U, $^{238}$U, and $^{232}$Th using monoenergetic photon // Nuclear Science and Engineering – 1980. – V. 73, N 2. – P. 153-163.

8. *Naik H., Goswami A., Kim G.N. et al* Mass-yield distributions of fission products from photofission of $^{232}$Th induced by 45 and 80 MeV bremsstrahlung // Physical Review C – 2012. – V. 86, Is. 5. 054607 [14 pages].

9. *Caldwell J.T., Dowdy E.J.* Experimental determination of photofission neutron multiplicities for eight isotopes in the mass range 232 ≤ A ≤ 239 // Nuclear Science and Engineering – 1975. – V. 56. – P. 179 - 187.

10. *Berman B.L., Caldwell J.T., Dowby E.I. et al.* Photofission and photoneutron cross sections and photofission neutron multiplicities for $^{233}$U, $^{234}$U, $^{237}$Np and $^{239}$Pu // Physical Review C – 1986. – V. 34, Is. 6. – P. 2201-2214.

11. *Verbeke J.M., Hagmann C., Wright D.* Simulation of neutron and gamma ray emission from fission and photofission // UCRL-AR-228518. Lawrence Livermore National Laboratory. January 17, 2014.
http://nuclear.llnl.gov/simulation/fission_v1.9/fission.pdf

12. WWW Table of radioactive isotopes. $^{241}$Am.

13. Desgrolard, Jenkovszky L., Lengyel A., Paccanoni F. Phys.Lett. B  - 1999.-v.459 – p.265-270.

14. *James F., Ross M.* Function minimization and error analysis. MINUIT D506. CERN Computer Centre Program library. – 1967. – P. 1- 47.

15. Photo-Nuclear Data. Average number of prompt neutrons released per fission event // Evaluated Nuclear Data File (ENDF). Database Version of April 13, 2015.
https://www-nds.iaea.org/exfor/endf.htm

16. *Watson S.J., Findlay D.J.S., Sené M.R.* Photofission and photoneutron measurements of $^{241}$Am between 5 and 10 MeV // Nuclear Physics A – 1992. – V. 548, Is. 3. – P. 365-373.